\documentclass[reprint,superscriptaddress,nofootinbib,nobibnotes, amsmath,amssymb, aps, prb, dvipsnames, floatfix]{revtex4-2}
\usepackage{graphicx}
\usepackage{dcolumn}  
\usepackage{bm}        
\usepackage{amssymb}   
\usepackage{verbatim}
\usepackage{todonotes}
\usepackage{xcolor}
\usepackage{url}

\newcommand{\qq}{\mathbf{q}}

\begin{document}

\widetext
\title{Superconductivity, Charge-Density-Waves, and Bipolarons in the Holstein model}
\author{B. Nosarzewski}
 \affiliation{Department of Physics, Stanford University, Stanford, CA 94305, USA }
 \affiliation{Stanford Institute for Materials and Energy Sciences (SIMES), SLAC National Accelerator Laboratory, Menlo Park, CA 94025, USA}

\author{E. W. Huang}%
 \affiliation{Department of Physics, Stanford University, Stanford, CA 94305, USA }
 \affiliation{Stanford Institute for Materials and Energy Sciences (SIMES), SLAC National Accelerator Laboratory, Menlo Park, CA 94025, USA}
 \affiliation{Department of Physics and Institute of Condensed Matter Theory, University of Illinois at Urbana-Champaign, Urbana, IL 61801, USA}
 
 \author{Philip M. Dee}
 \affiliation{Department of Physics and Astronomy, University of Tennessee, Knoxville, TN 37996, USA }

\author{I. Esterlis}
\affiliation{Department of Physics, Stanford University, Stanford, CA 94305, USA }
\affiliation{Department of Physics, Harvard University, Cambridge, MA 02138, USA }

\author{B. Moritz}%
 \affiliation{Stanford Institute for Materials and Energy Sciences (SIMES), SLAC National Accelerator Laboratory, Menlo Park, CA 94025, USA}
 
\author{S. A. Kivelson}
\affiliation{Department of Physics, Stanford University, Stanford, CA 94305, USA }

\author{S. Johnston}
 \affiliation{Department of Physics and Astronomy, University of Tennessee, Knoxville, TN 37996, USA }

\author{T. P. Devereaux}%
 \affiliation{Stanford Institute for Materials and Energy Sciences (SIMES), SLAC National Accelerator Laboratory, Menlo Park, CA 94025, USA}
 \affiliation{Department of Materials Science and Engineering, Stanford University, Stanford, California 94305, USA}
 \affiliation{Geballe Laboratory for Advanced Materials, Stanford University, Stanford, CA 94305, USA}

\date{\today}

\begin{abstract}
The electron-phonon ($e$-ph) interaction remains of great interest in condensed matter physics and plays a vital role in realizing superconductors, charge-density-waves (CDW), and polarons. We study the two-dimensional Holstein model for $e$-ph coupling using determinant quantum Monte Carlo across a wide range of its phase diagram as a function of temperature, electron density, dimensionless $e$-ph coupling strength, and the adiabatic ratio of the phonon frequency to the Fermi energy. We describe the behavior of the CDW correlations, the competition between superconducting and CDW orders and polaron formation, the optimal conditions for superconductivity, and the transition from the weak-coupling regime to the strong-coupling regime. Superconductivity is optimized at intermediate $e$-ph coupling strength and intermediate electron density, and the superconducting correlations increase monotonically with phonon frequency. The global maximum for superconductivity in the Holstein model occurs at large phonon frequency, the limit where an attractive Hubbard model effectively describes the physics.
\end{abstract}

\maketitle

\section{Introduction}
Electron-phonon ($e$-ph) coupling is ubiquitous in quantum materials and leads to superconductivity (SC), charge-density-wave (CDW) order, and the formation of polarons.\cite{BCS, Scalapino, CDW, polaron, Marsiglio, Johnston} The properties of many-body quantum systems with $e$-ph interactions can be calculated perturbatively in certain limits. For example, in the limit of weak coupling and small phonon frequency, these systems can be described by Migdal-Eliashberg theory.\cite{Migdal, Eliashberg, Engelsberg1963, MarsiglioPRB1990, DeePRB2019,  NosarzewskiPRB2021} Conversely, the strong coupling limit can be treated using the Lang-Firsov transformation, which sets up perturbation theory around the polaronic state.\cite{LangFirsov, FreericksStrongCoupling, Giustino} Many materials, however, fall into the intermediate coupling regime for which no general analytic solution exists. In this context, it is especially interesting to consider the intermediate coupling regime because the superconducting transition temperature $T_c$ tends to zero in both the weak-coupling and strong-coupling limits, implying that it is maximized somewhere in between. 

Competition with lattice instabilities arising from CDW order or polaron formation has important consequences for superconductivity.\cite{MarsiglioPRB1990, DeePRB2019, LiEPL2015, ScalettarCompetition, Noack, Vekic, Meyer, Capone, FreericksDMFT, Benedetti, Esterlis} Even for materials where strong electron correlations may play a dominant role, as in unconventional superconductors such as the cuprates, the presence of significant $e$-ph coupling has been established through the presence of strong renormalizations in measurements of bandstructure, \cite{LanzaraNature,CukPRL2004,PlumbPRL2010,Lee} phonon lineshape in Raman measurements,\cite{Devereaux, Farina, Zhang} neutron scattering,\cite{Pintschovius}  and the observation of unconventional isotope effects.\cite{CrawfordPRB1990, ChenPNAS2007} The interplay between SC and CDW order mediated by $e$-ph coupling could be an important effect that limits the superconducting transition temperature in the cuprates,\cite{Keimer, Chang} as well as in other materials with SC and CDW phases such as $\textrm{2H-TaS}_2$,\cite{Pablo} the bismuthates BaPb$_{1-x}$Bi$_{x}$O$_3$ and Ba$_{1-x}$K$_x$BiO$_3$,\cite{Sleight,Fisher}  tri-tellurides,\cite{Zocco} and pnictides.\cite{LeePreprint2021} This physics may also be relevant to the A15 compounds, which are close to a lattice instability that is not a CDW but still another form of charge order.\cite{StewartReview2015} 

The properties of $e$-ph mediated superconductors are often well described by Migdal-Eliashberg theory, which is nominally valid when $\lambda\frac{\hbar \Omega}{E_{\text{F}}} \ll 1$. (Here, $\lambda$ is the dimensionless $e$-ph coupling strength, $\hbar\Omega$ is the phonon energy, and $E_{\text{F}}$ is the Fermi energy.\cite{Migdal, Eliashberg}) But for several novel superconductors including \textit{n}-type $\textrm{SrTiO}_3$ \cite{STO}, monolayer $\textrm{FeSe}$ on $\textrm{SrTiO}_3$,\cite{FESTO} the fullerides,\cite{Grimaldi} and lightly doped oxides,\cite{Gunnarsson}, the phonon frequency is large compared to the Fermi energy. To fully understand these systems, one must be able to accurately compute the properties of $e$-ph systems in the anti-adiabatic limit ($\hbar\Omega > E_{\text{F}}$) and assess the accuracy of Migdal-Eliashberg theory in this regime. \cite{Esterlis}

\begin{figure*}
\centering
\includegraphics[scale=0.32]{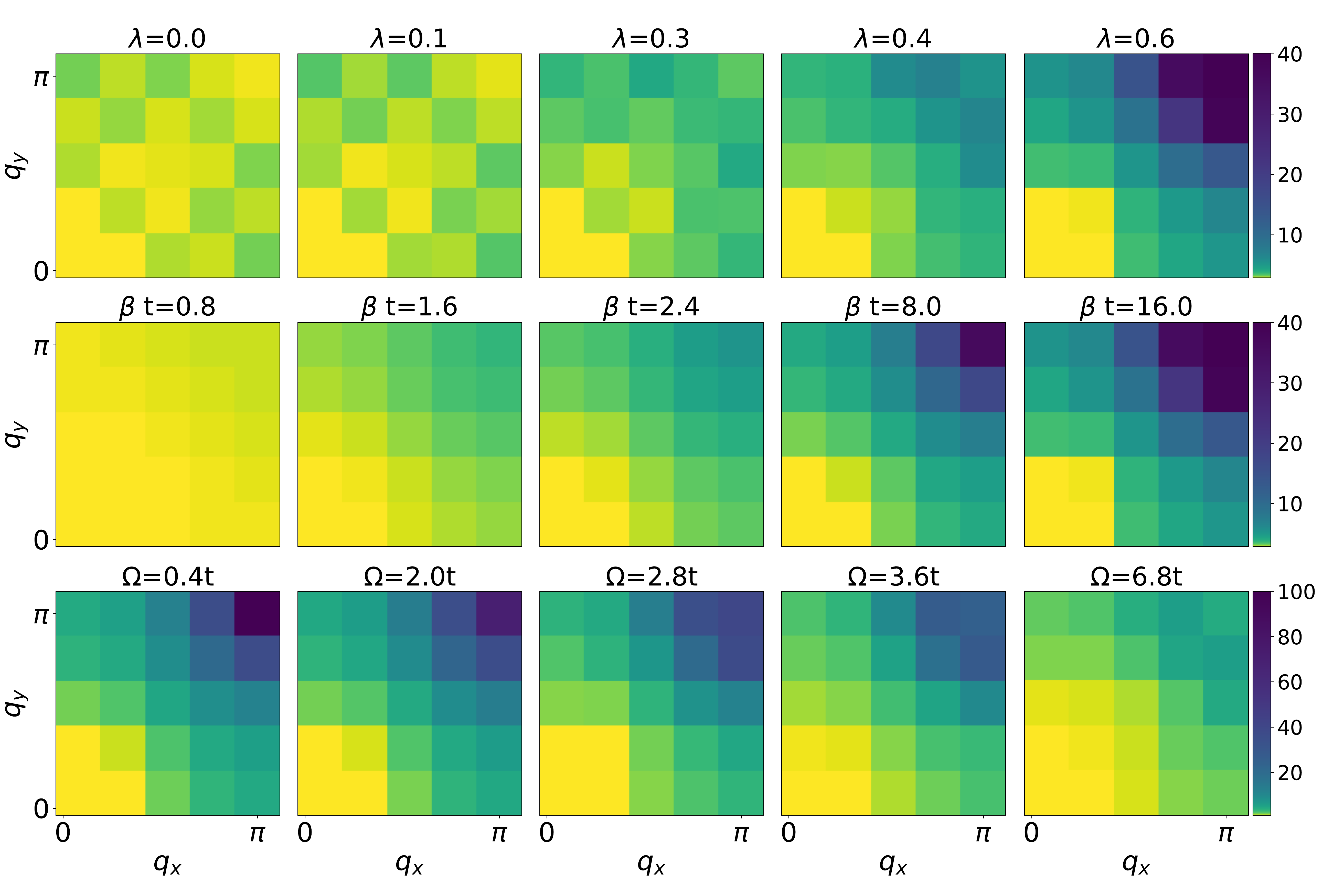}
\caption[width=\textwidth]{All panels show the momentum dependence of the CDW susceptibility for an $8 \times 8$ lattice at a filling of $\langle n \rangle = 0.6$. The first row shows the behavior of the susceptibility as a function of $\lambda$ also for a fixed phonon frequency of $\Omega=2.8t$ and fixed temperature of $\beta t=16$. The second row shows the temperature dependence of the susceptibility for a phonon frequency $\Omega=2.8t$ and an $e$-ph coupling $\lambda=0.6$. The third row shows the $\Omega$ dependence of the susceptibility for a fixed $e$-ph coupling $\lambda=0.6$ and for a fixed temperature of $\beta t=16$.}
\label{wavevector}
\end{figure*}

The Holstein model is a paradigmatic model of $e$-ph coupling, consisting of electrons locally coupled to a single dispersionless optical phonon branch.\cite{Holstein} Despite its simplicity, the model contains the essential physics of $e$-ph coupled systems, including SC and CDW orders and polaron formation. Here, we use determinant Quantum Monte Carlo (DQMC) to obtain numerically exact results for the SC and CDW susceptibilities of the two-dimensional Holstein model, which allows for competition between these orders and additional polaronic effects.  While several previous studies of the Holstein model have investigated similar questions using non-perturbative methods,\cite{ScalettarCompetition, Noack, Vekic, Meyer, Capone, FreericksDMFT, Benedetti, Dee2020, owen_scalettar} most of them (except for the most recent \cite{owen_scalettar}) were limited to narrow regions of the phase diagram or were performed in infinite dimensions using dynamical mean-field theory. Motivated by this, we have carried out a comprehensive study across a broad region of the phase diagram spanning weak, intermediate, and strong $e$-ph coupling, various phonon frequencies, and a wide range of doping. 

We find that superconductivity is generally optimized at intermediate values of the $e$-ph coupling strength and electron density. We also find that the strength of the SC correlations increases monotonically with phonon frequency until saturating in the extreme anti-adiabatic limit, where the model is equivalent to the attractive Hubbard model. Since our results are obtained in the absence of any Coulomb repulsion, they should be viewed as reflecting limits on the Holstein interaction in an idealized setting. Additional interactions, especially the Coulomb interaction, that are omnipresent in real materials, will place further constraints on the superconducting $T_c$ ultimately realized. 

\section{Model and Methods}
\subsection{Models}
The Hamiltonian for the Holstein model\cite{Holstein} linearly couples the electron density at each lattice site to the displacement of an independent harmonic oscillator at that site and is given by 
\begin{equation}
\begin{split}
H = &-t\sum_{\langle ij \rangle \sigma}(c^\dagger_{i\sigma} c^{\phantom{\dagger}}_{j\sigma} + \textrm{H.c.})- \mu \sum_{i\sigma}c^\dagger_{i\sigma}c^{\phantom{\dagger}}_{i\sigma}  \\ &+ \sum_i \left( \frac{p_i^2}{2M} + \frac{1}{2}M\Omega^2x_i^2 \right) - g \sum_i n_i x_i. 
\end{split}
\end{equation}
Here, $c^\dagger_{i\sigma}$ creates an electron on site $i$ with spin $\sigma=\uparrow,\downarrow$, $\langle \cdot \rangle$ denotes a sum over nearest neighbor sites, $t$ is the nearest-neighbor hopping integral, $n_i = c^\dagger_{i\uparrow}c^{\phantom{0}}_{i\uparrow}+c^\dagger_{i\downarrow} c^{\phantom{0}}_{i\downarrow}$ is the local electronic density, $\mu$ is the chemical potential, $x_i$ and $p_i$ are the position and momentum operators of independent harmonic oscillators with mass $M$ and frequency $\Omega$, and $g$ is the $e$-ph coupling constant. Throughout this work, we take units in which $\hbar=k_\text{B}=M=a=t=1$, and consider the system on a two-dimensional square lattice. The dimensionless parameters for the Holstein model are the dimensionless $e$-ph coupling $\lambda$, the adiabatic ratio $\Omega/E_{\text{F}}$, and the average electron density $\langle n \rangle$. Two common definitions of the dimensionless $e$-ph coupling 
can be found in the literature. The first is $\lambda = \frac{g^2}{M\Omega^2 W}$, where $W=8t$ is the bandwidth. The second is $\lambda_0 = \frac{g^2 N(0)}{M\Omega^2}$, where $N(0)$ is the density of states at the Fermi level. 
The former is more commonly used in QMC calculations while the latter frequently appears in the context of Migdal-Eliashberg calculations. We will discuss our results in terms of both definitions to facilitate connections to previous works using either one.  

\subsection{Methods}
We study the Holstein model using DQMC, which is a non-perturbative method that stochastically evaluates finite temperature expectation values in imaginary time.\cite{WhiteDQMC} Details of the DQMC algorithm, including the explanation of local and global phonon field updates, can be found in Ref. \onlinecite{DQMC}. DQMC is sign problem-free for the Holstein model, but it suffers from long phonon autocorrelation times in the regimes of large $e$-ph coupling, low phonon frequency, or low temperature. In our simulations, we access temperatures down to $T=\beta^{-1}=t/16$ for $8\times8$ lattices with periodic boundary conditions.  A typical Markov chain for importance sampling in the Monte Carlo process consists of approximately $20k$ warm-up sweeps for equilibration and $100k$ measurement sweeps, which at the lowest temperatures (largest $\beta$) takes approximately $8$ hours.  As an estimate for the computational cost at the lowest temperature for most of the data presented here, 20 Markov chains were run for 9 values of the phonon frequency $\Omega$, 20 electron densities $\left< n \right>$ between 0 and 1, and 12 values of $\lambda$ between 0 and 0.6 spaced by $\Delta\lambda=0.05$, requiring approximately $350k$ CPU hours.

The SC and CDW correlations in the system can be accessed by measuring their 
respective susceptibilities. The SC pair-field susceptibility is defined as
\begin{equation}
\chi^{\phantom{\dagger}}_\text{SC} = \int_0^\beta d\tau \langle \Delta(\tau) \Delta^\dagger(0) \rangle,
\end{equation}
where 
\begin{equation}
\Delta^\dagger = \frac{1}{L} \sum_i c^\dagger_{i\uparrow} c^\dagger_{i\downarrow}
\end{equation}
and $L=8$ is the linear size of the system. The charge susceptibility is  
\begin{equation}
\chi^{\phantom{\dagger}}_\text{CDW}(\textbf{q}) = \int_0^\beta d\tau \langle \rho_\textbf{q}(\tau) \rho^\dagger_\textbf{q}(0) \rangle,
\end{equation}
where 
\begin{equation}
\rho^\dagger_\textbf{q} = \frac{1}{L} \sum_{i\sigma} e^{\mathrm{i}\textbf{q} \cdot \textbf{R}_i} c^\dagger_{i\sigma}c^{\phantom{0}}_{i\sigma}.
\end{equation} 
In the thermodynamic limit, the temperature at which the charge and superconducting pair-field susceptibilities diverge determines the CDW and SC transition temperatures, respectively. While we do not access sufficiently low temperatures and sufficiently large lattice sizes to observe transitions to either a SC or a CDW phase, we do access temperatures low enough to identify a significant growth of the corresponding susceptibilities. We can, therefore, determine the dominant ordering tendencies of the system. We also note that our choice for the operator $\Delta$ assumes that pairing occurs between electrons in the $s$-wave channel. This definition will have finite overlap with the superconducting state that forms when the quasiparticles are polarons, provided that the polaron's quasiparticle weight is non-zero. 
 
 \begin{figure}
\includegraphics[width=\columnwidth]{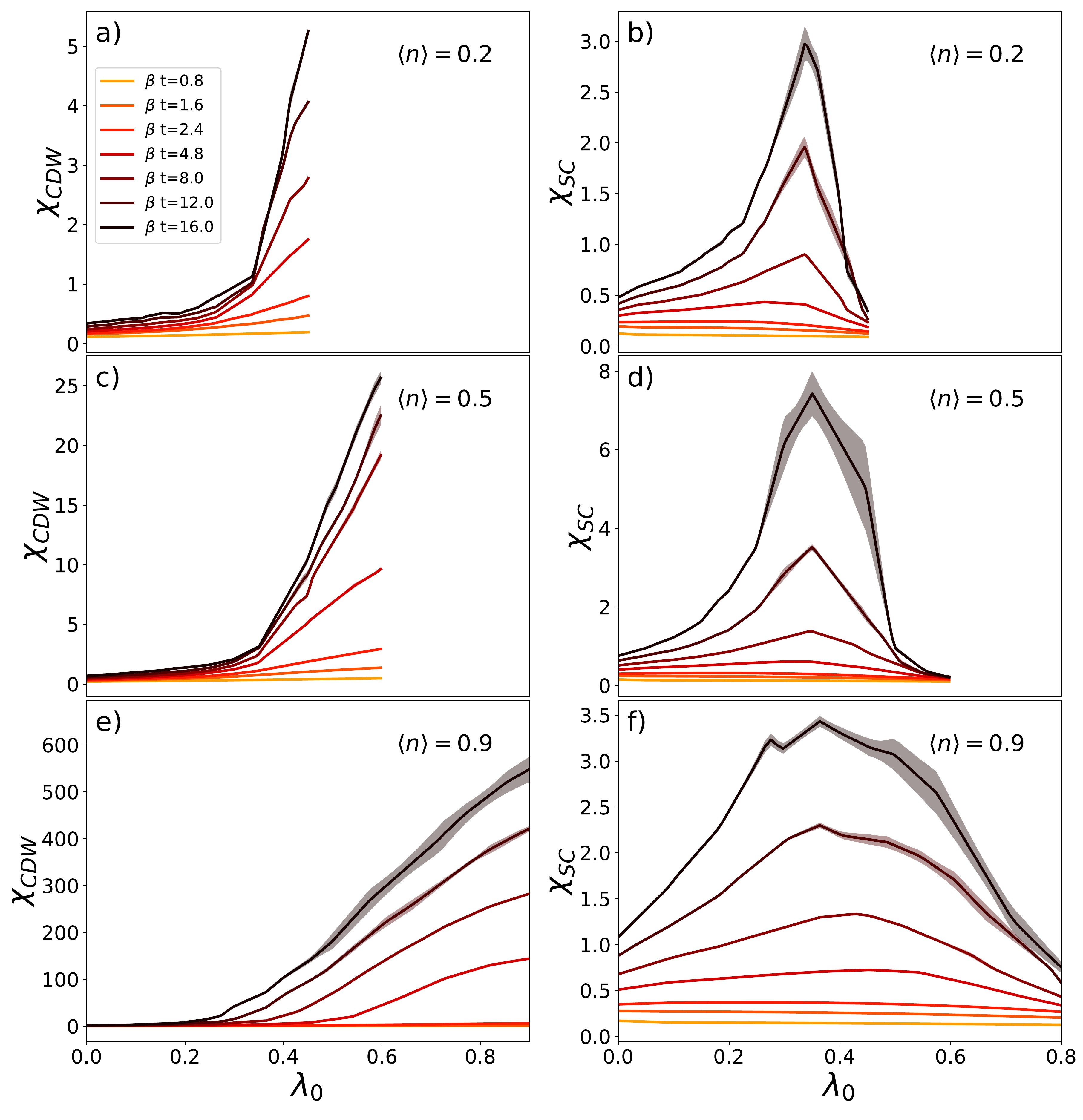}
\caption{All panels are DQMC results for an $8 \times 8$ lattice for a phonon frequency of $\Omega=2t$. Each panel shows the SC or maximum of the CDW susceptibility as a function of $\lambda_0$ for several different temperatures. Shaded regions represent the standard error of the data. a,c,e) Show the charge-density-wave susceptibility for fillings of $\langle n \rangle=0.2,0.5,0.9$ respectively. b,d,f) Show the maximum of the superconducting susceptibilities for fillings of $\langle n \rangle=0.2,0.5,0.9$ respectively. }
\label{lambda}
\end{figure}

\begin{figure}
\includegraphics[width=\columnwidth]{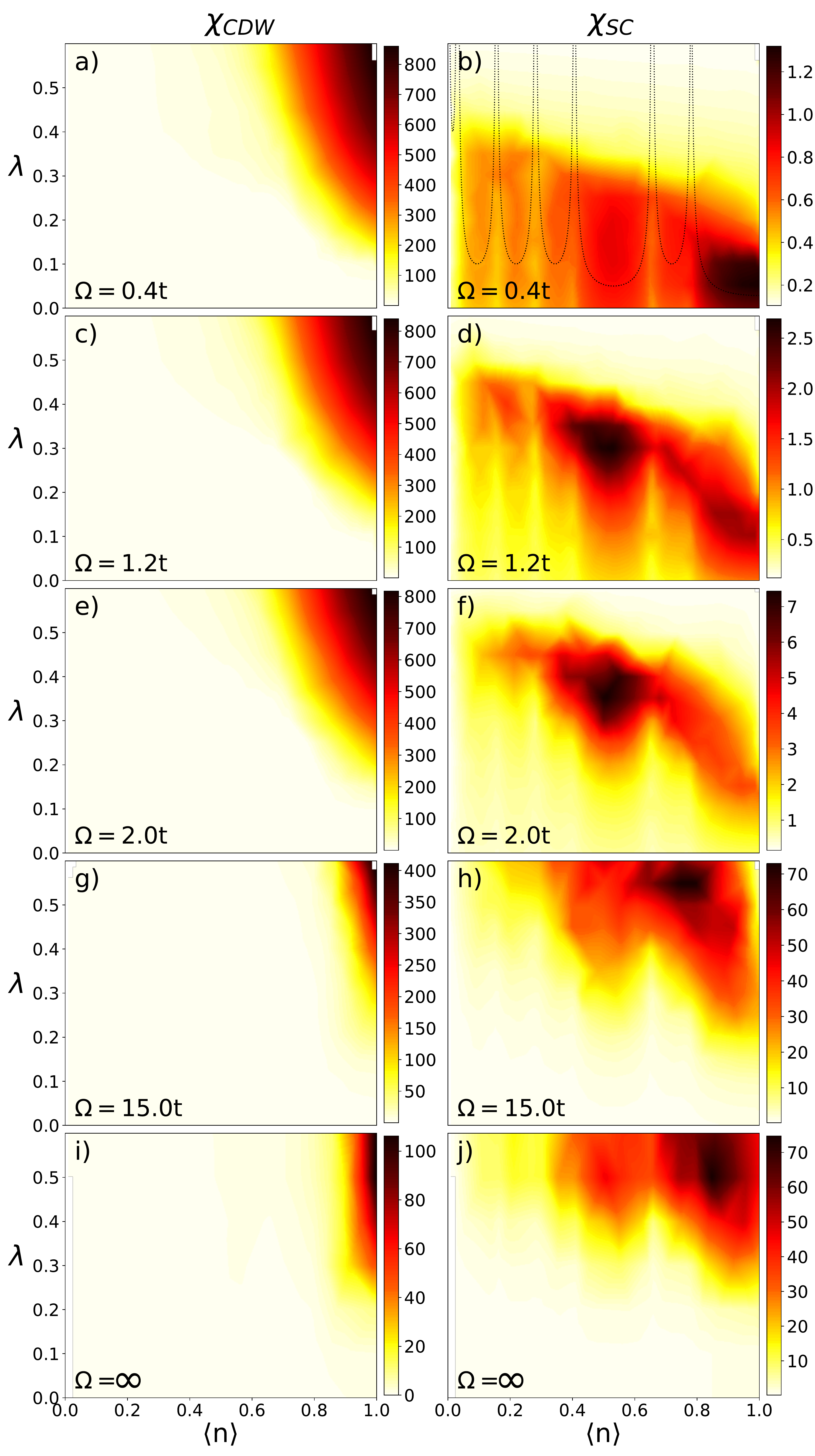}
\caption{All panels are DQMC results for an $8 \times 8$ lattice for $\beta t=16$. Each panel is for a different phonon frequency determined across 12 values of $\lambda$, equally spaced by $\Delta\lambda = 0.05$ between $0$ and $0.6$,  and 20 values of the filling $\left< n \right>$, spaced approximately equally between $0$ and $1$. The ``heat map" plots are obtained by linearly interpolating the susceptibility onto a regular two dimensional grid using Python's \texttt{scipy.interpolate.griddata function\cite{scipy-interpolate}}. Divergences in the dotted line in panel b) are calculated by taking the derivative of the filling with respect to the chemical potential for the non-interacting bandstructure. Infinite phonon frequency implies simulations of the negative-U Hubbard model. a,c,e,g,i) Maximum (over all momenta) of the charge-density-wave susceptibility for phonon frequencies $\Omega = 0.4t,\ 2.0t,\ 6.8t,\ 15.0t,\ \infty$. b,d,f,h,j) Superconducting susceptibility for the same phonon frequencies. }
\label{doping_lambda}
\end{figure}

\begin{figure}
	\includegraphics[scale=0.42]{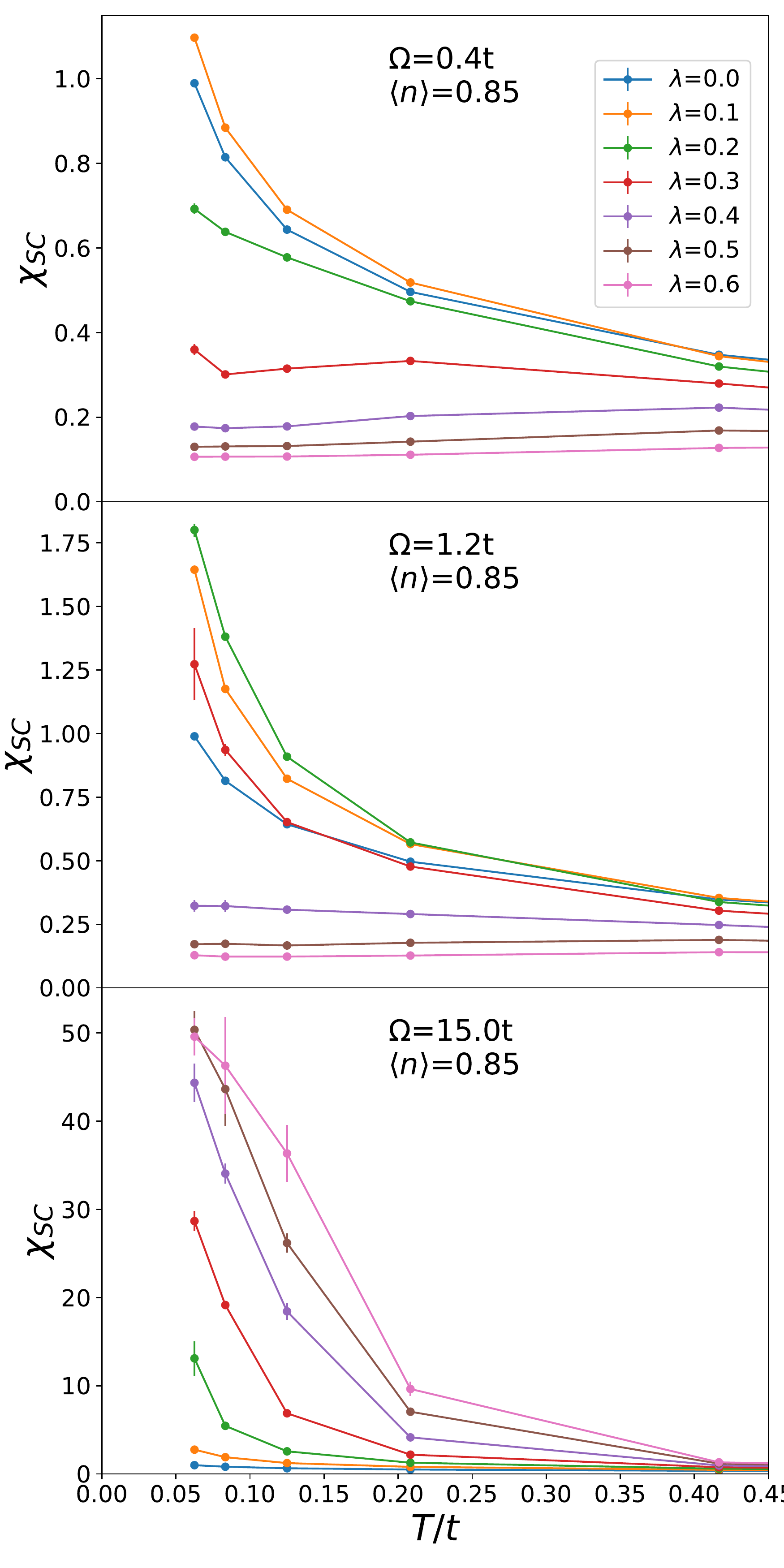}
	\caption{All panels are DQMC results for an $8 \times 8$ lattice. Each panel shows the SC susceptibility for different values of the phonon frequency as a function of temperature and $\lambda$ for a filling of $\langle n \rangle = 0.85$.}
	\label{T}
\end{figure}

\section{Results}
\subsection{Charge-density-wave susceptibility}

We first discuss the momentum dependence of the CDW susceptibility $\chi^{\phantom{\dagger}}_\text{CDW}(\textbf{q})$ as a function of $\lambda_0$, temperature, and $\Omega$. To this end, Fig.~\ref{wavevector} plots the momentum dependence of the CDW susceptibility as a function of $\lambda_0$ at fixed filling $\langle n \rangle=0.6$.

In the limit of $\lambda_0\rightarrow 0$, the charge response is determined by the Lindhard function and governed mainly by the Fermi surface's shape. At weak $e$-ph coupling, the CDW ordering wavevector remains tied to the shape of the Fermi surface and closely resembles the Lindhard response, as seen in the first two panels of the first row of Fig.~\ref{wavevector}.  

As $\lambda$ increases, a cross-over occurs to the strong coupling limit. Here, the susceptibility becomes strongly peaked at ${\bf q} = (\pi,\pi)$ and disconnected from the shape of the Fermi surface, even though the filling $\langle n \rangle=0.6$ is far from half-filling. In other words, the CDW correlations are dominated by $(\pi,\pi)$ ordering tendencies at low temperatures in the strong coupling limit, even away from half-filling, where a weak-coupling picture would predict an incommensurate CDW ordering wavevector~\cite{DeePRB2019}. This occurs even though the density is not naturally high enough to place two electrons on every other site in a checkerboard pattern corresponding to the $(\pi,\pi)$ wavevector. Generally, the ordering wavevector will depend on the details of the model, such as the electronic and phononic bandstructure and the momentum dependence of the $e$-ph coupling.

The expected behavior in the strong coupling limit can be understood using a bipolaron picture and a Lang-Firsov transformation, which shifts the equilibrium position of each oscillator to $-g n_i / M \Omega^2$ and creates an effective attractive interaction between electrons that encourages double occupancy at each site. 
Such a double occupation with an associated lattice distortion is known as a bipolaron and has a binding energy of $-g^2 / M \Omega^2 = -\lambda W$.\cite{FreericksStrongCoupling} In the strong coupling limit at higher temperatures, as shown in the second row of Fig.~\ref{wavevector}, the susceptibility is still peaked at $(\pi,\pi)$. Still, the peak is broader, which is a signature of bipolaron formation despite the lack of long-range CDW order. In this case, the peak in $\chi_\text{CDW}(\pi,\pi)$ is due to the presence of fluctuating short-range charge 
order. Bipolaron formation is a generic feature of $e$-ph systems in the strong-coupling regime, 
independent of the model details.\cite{esterlis_pseudogap, Li2020} 

The dominant wavevector in $\chi_\text{CDW}({\bf q})$ also depends on the phonon frequency, as shown in the third row of Fig.~\ref{wavevector}. Even with $\lambda$ fixed, large phonon energies produce weaker $(\pi,\pi)$ CDW correlations, as evidenced by the decrease in magnitude and the shift in ordering wavevector of the susceptibility away from $(\pi,\pi)$. The tendency towards CDW order is weaker in the large phonon frequency (anti-adiabatic) limit, as the lattice responds more quickly to electronic hopping. The lattice deformations that form potential wells trapping electrons in place become weaker as a result. 

With the general behavior of the momentum dependence of $\chi^{\phantom{0}}_\text{CDW}(\textbf{q})$ established, we now focus on the magnitude of the CDW susceptibility as a function of temperature, $\lambda$, $\Omega$, and filling. Since the CDW susceptibility is a function of wavevector $\qq$, we choose the $\qq = \qq_\text{max}$ at which the susceptibility is maximized when reporting the magnitude. In Fig.~\ref{lambda}(a,c,e), the magnitude (maximum) of the $\chi_\text{CDW}(\qq_\text{max})$ is shown as a function of $\lambda_0$ for different temperatures and fillings. The magnitude of the CDW susceptibility is small and relatively featureless at high temperature but exhibits a rapid increase at low temperatures as $\lambda_0$ increases. In general, we find that the CDW correlations can be significant, even at dilute concentrations [Fig.~\ref{lambda}a)]; however, we also observe a dramatic increase in $\chi^{\phantom{0}}_\text{CDW}(\textbf{q}_\text{max})$ as 
the density approaches half-filling [Fig.~\ref{lambda}e)]. This behavior may reflect that the Fermi surface provides better nesting for the $(\pi,\pi)$ wavevector, which naturally enhances the CDW tendencies. We also generally find that the CDW susceptibility exhibits faster growth with increasing filling. Nevertheless, at strong enough coupling and low temperature, there is a tendency towards CDW order and/or phase separation even at low filling~\cite{esterlis_pseudogap, owen_scalettar}.  

To demonstrate the effect of phonon frequency, and to better visualize the behavior of the susceptibility smoothly as a function of filling and $\lambda_0$, we plot the magnitude of $\chi_\text{CDW}(\qq_\text{max})$ at the lowest temperature ($\beta t=16$) as a function of both filling and $\lambda_0$ in Fig.~\ref{doping_lambda}(a,c,e,g,i). Increasing the phonon frequency suppresses the CDW correlations, which is apparent from the decrease in the magnitude of $\chi_\text{CDW}(\qq_\text{max})$ and its weakened influence in regions of lower $\lambda_0$ and away from half-filling.    

In summary, the CDW tendencies are enhanced for values of large $\lambda$, near half-filling, and at small phonon frequency. Next, we will discuss the behavior of the SC correlations in the context of the CDW correlations. We will see that SC order is overshadowed by the CDW order in the regimes where the CDW correlations are strongest. In other words, the SC order is confined to intermediate values of $\lambda$ and intermediate filling and becomes stronger with increasing phonon frequency.

\section{Superconducting susceptibility}

The weak-coupling limit is associated with the existence of a Fermi surface with well-defined quasiparticles, a picture that breaks down at large $\lambda$ due to (bi)polaron formation and/or lattice instability. In the weak-coupling, adiabatic limit, a BCS superconducting state is formed at low temperature, where the transition temperature is expected to behave as $T_c \sim \Omega e^{-\frac{1}{\lambda}}$.\cite{BCS} The rate of growth of the SC susceptibility with temperature in Fig.~\ref{lambda}(b,d,f) shows an initial increase with $\lambda_0$ consistent with this expectation.  The suppression of the SC susceptibility at large values of $\lambda_0$ for three different values of the electron density, as shown in Fig.~\ref{lambda}(b,d,f), indicates a breakdown of the weak-coupling BCS prediction. The region where this occurs corresponds precisely to the region where the CDW correlations begin to grow rapidly with temperature, as seen in Fig.~\ref{lambda}(a,c,e). 

From Fig.~\ref{lambda}, we infer that the suppression of superconductivity in the large $\lambda$ regime occurs as a result of competition with CDW order and a tendency toward bipolaron formation, which in turn generates short-range ($\pi,\pi$) CDW correlations, even in the absence of charge order. This competition occurs even at a low filling (see $\langle n \rangle=0.2$) well away from the strongest CDW tendencies. Comparing the SC susceptibility at three selected fillings shown in Fig.~\ref{lambda}, the strongest rate of increase in $\chi_\text{SC}$ occurs at intermediate filling ($\langle n \rangle=0.5$). A low filling is not favorable to superconductivity as there are fewer electrons available to form a condensate. Conversely, a strong tendency towards $(\pi,\pi)$ CDW order dominates the tendency towards SC order near half-filling. Moreover, we observe a peak in the SC susceptibility around $\lambda_0 \approx 0.4$ across all three fillings. We remark that $\lambda_0$, defined from the bare microscopic parameters in the Hamiltonian, is in general different from the physical coupling strength, $\lambda_\text{phys}$, extracted from, e.g., tunneling experiments \cite{allen1982}. In general, phonon softening will tend to increase the coupling strength, so that $\lambda_\text{phys} > \lambda_0$. In numerical calculations $\lambda_\text{phys}$ may be extracted from the fully dressed phonon propagator \cite{Esterlis,NosarzewskiPRB2021}. We estimate that $\lambda_0 \approx 0.4$ corresponds roughly to $\lambda_\text{phys} \approx 1-2$, depending on other parameters, which is a range compatible with coupling strengths known from the study of strongly-coupled electron-phonon superconductors. 

These observations suggest that the optimal regime for superconductivity at this particular phonon frequency ($\Omega=2.8t$) occurs at an intermediate filling and $\lambda$. As the phonon frequency increases, as shown in Fig.~\ref{doping_lambda}(b,d,f,h,j), the strength of the CDW susceptibility decreases and the magnitude of the SC susceptibility increases and the optimal filling shifts closer to half-filling. For the largest phonon frequency used in our simulation ($\Omega=15t$), the SC and CDW susceptibilities shown in Fig.~\ref{doping_lambda}(g,h) quantitatively and qualitatively approach DQMC results for the attractive Hubbard model shown in Fig.~\ref{doping_lambda}(i,j). In the limit that the phonon frequency approaches infinity, the Holstein model can be mapped onto an attractive Hubbard model with an on-site attraction $U=-\lambda W$, and the SC tendency is optimized at an intermediate filling.\cite{hubbard} Note that non-smooth features in the susceptibilities arise from the finite-size effects associated with the $8 \times 8$ cluster used for the simulation. The divergences of the dotted line in Fig.~\ref{doping_lambda}b indicate special fillings due to the discrete sampling of momentum space. They roughly correspond to the fillings at which finite size effects are expected to be more noticeable. 
  
Figure~\ref{T} explicitly shows $\chi_\text{SC}$ as a function of temperature. As described above, the result generally indicates that SC is optimized at an intermediate $\lambda$. Moreover, the rate of increase of the SC susceptibility grows with increasing phonon frequency, and the optimal value of $\lambda$ moves towards larger coupling strengths. 
 
\begin{figure}
	\includegraphics[width=\columnwidth]{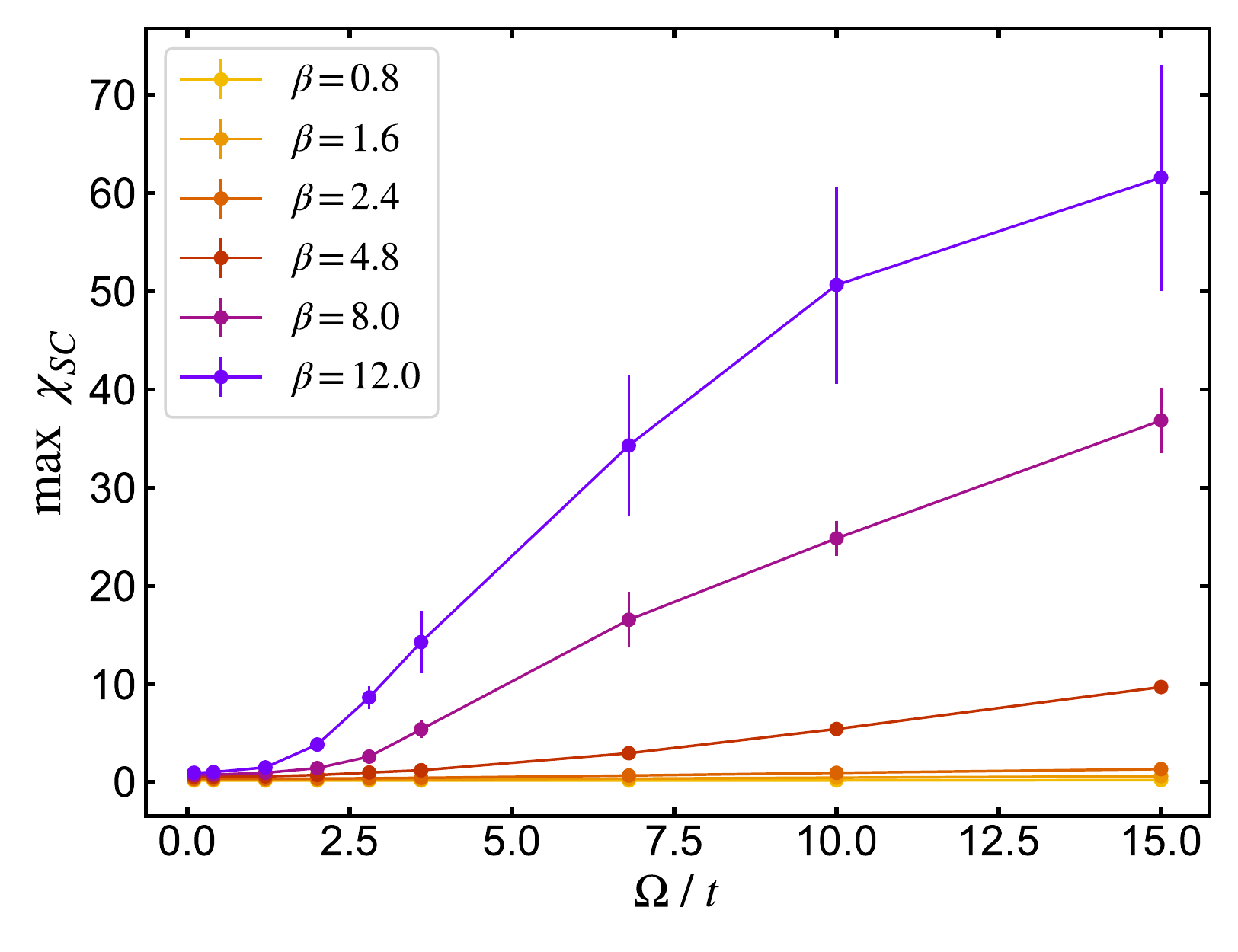}
	\caption{Maximum of superconducting susceptibility across electron density and $\lambda$ as a function of phonon frequency and temperature for an $8 \times 8$ lattice. }
	\label{max_xsc}
\end{figure}
 
We now make a statement about the behavior of the global maximum of the superconducting $T_c$ in the Holstein model by investigating the maximum value of the superconducting susceptibility across a wide range of electron densities and $e$-ph coupling strengths.  Figure~\ref{max_xsc} shows that the maximum value of $\chi_\text{SC}$ increases monotonically as a function of $\Omega$ all the way to $\Omega=15t$. We previously showed that the DQMC results at $\Omega=15t$ closely resemble the infinite phonon frequency limit described by the attractive Hubbard model, so the monotonic increase in $\chi_\text{SC}$ with $\Omega$ effectively extends to infinite phonon frequencies. From its temperature dependence, it is further apparent that the rate of $\chi_\text{SC}$'s growth increases as a function of temperature, implying that the superconducting transition temperature also increases monotonically as a function of phonon frequency. Ref.~\onlinecite{esterlistc} has argued that the upper bound on the superconducting $T_c$ in the Holstein model is set by $T_c \approx 0.1 \Omega$ in the adiabatic limit. In the opposite limit of infinite phonon frequency, Ref.~\onlinecite{moreo} has shown that the maximum superconducting transition temperature is $T_c \approx 0.2 t$ based on studies of the attractive Hubbard model. The known behavior in these two limits, together with our 
results in Fig.~~\ref{max_xsc}, then imply that the maximal $T_c$ in the Holstein model must increase monotonically as a function of $\Omega$, behaving linearly at small $\Omega$ (based on the predictions of BCS and ME theory) and then plateauing at the maximal $T_c \approx 0.2t$ of the attractive Hubbard model.

It would be interesting in future work to study the transition between the linear regime in the adiabatic limit and the plateau in the anti-adiabatic limit. This transition will occur around $\Omega \approx 2 t$, a regime where maximal $T_c$ is likely accessible in DQMC simulations. We expect that optimizing the details of the bandstructure (such as including the effect of next-nearest neighbor hopping) will not qualitatively change these conclusions because the optimal densities for superconductivity occur away from half-filling, where the particular shape of the Fermi surface avoids any special nesting conditions. So, one should expect the suppression of SC at strong coupling due to competition with CDW order or bipolaron formation, regardless of the details of the bandstructure. 

Finally, we remark that we have neglected the Coulomb repulsion entirely in the current study. In the adiabatic limit of small phonon frequency, we expect our conclusions to remain qualitatively unchanged, as the dimensionless Coulomb repulsion $\mu_C$ is renormalized down $\mu_C \to \mu_C^* < \mu_C$ in this limit, and the pairing tendencies of the system will be determined by the combination $\lambda_0 - \mu_C^* \lesssim \lambda_0$~\cite{allen1982}.  Retardation becomes less effective in suppressing the Coulomb repulsion in the limit of large phonon frequency, however. In the particular case that the repulsion is modeled by an on-site Hubbard $U>0$, the behavior in the antiadiabatic limit $\Omega \to \infty$ will be that of an effective Hubbard model with $U_\text{eff} = U -  \lambda_0/N(0)$. In general, $U_\text{eff} > 0$, so one expects the tendency toward superconductivity to be suppressed in this regime. These considerations suggest superconductivity is optimized at an intermediate value of the phonon frequency once Coulomb repulsion is taken into account. (An interested reader is referred to Refs.~\onlinecite{DQMC, BergerPRB1995, CaponePRB2006, Bauer, Weber, Mendl2017, Karakuzu, Han2020, Costa2020} and references therein for further discussion of the single-band Hubbard-Holstein model.)

\begin{figure}
	\includegraphics[width=\columnwidth]{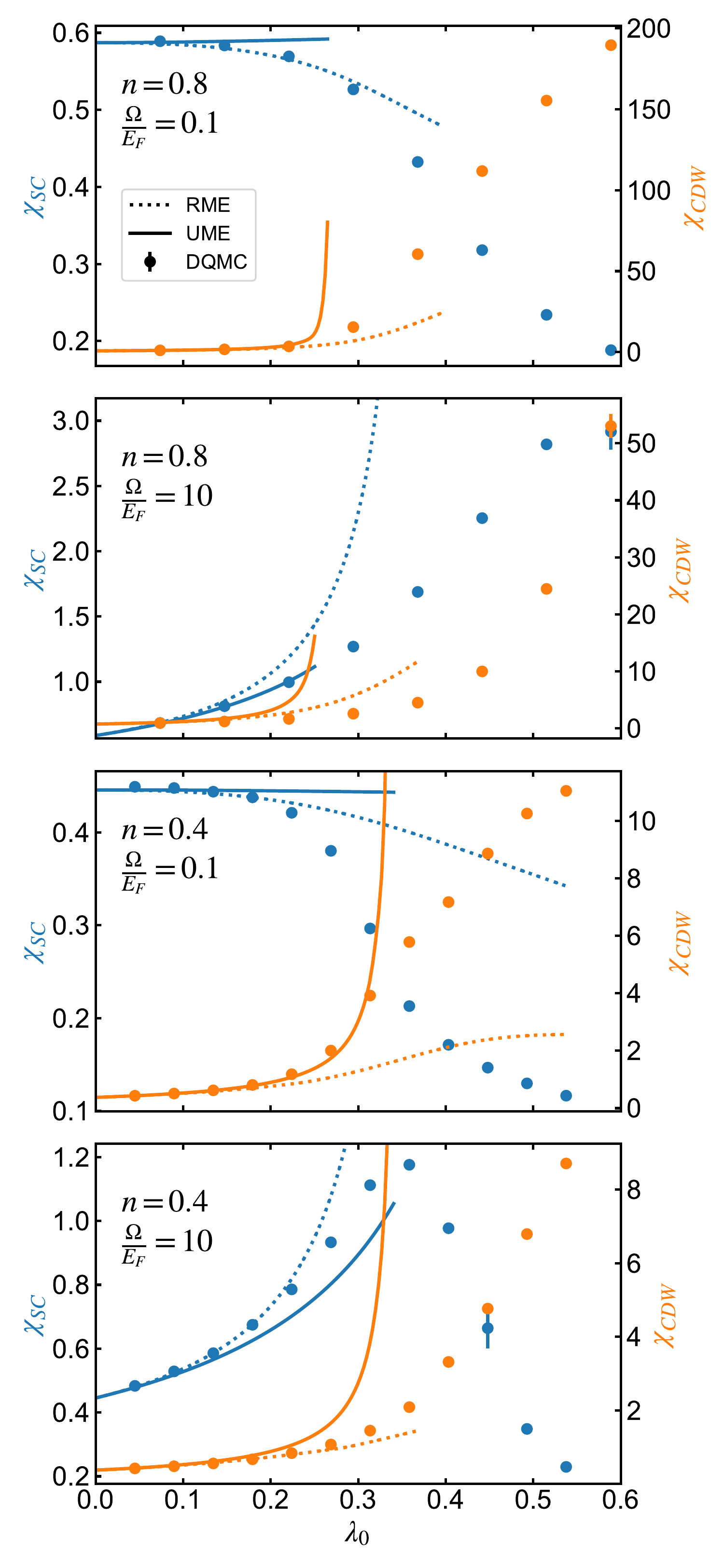}
	\caption{\label{fig3} Comparison of superconducting and charge-density-wave (CDW) susceptibilities from Migdal-Eliashberg (ME) and determinant quantum Monte Carlo (DQMC) for an $8 \times 8$ lattice and $\beta t=8$. The value of the $\chi_\text{CDW}(\qq)$ represents its value at the wave-vector for which it is maximized. Solid lines represent unrenormalized ME theory, dotted lines represent renormalized ME theory, and points represent DQMC data. Superconducting (SC) susceptibilities are shown in blue and SC susceptibilities are shown in orange. The relation between $\lambda_0$ and $\lambda$ is $\lambda_0 = N(0) W \lambda$. For $\langle n \rangle=0.4$, $\lambda_0 = 0.90 \lambda$ and for $\langle n \rangle=0.8$, $\lambda_0 = 1.5 \lambda$. }
	\label{migdal}
\end{figure}

\section{Comparison with Migdal-Eliashberg theory}

Migdal-Eliashberg (ME) theory provides a foundation for our understanding of conventional superconductors.\cite{Migdal, Eliashberg} As we have discussed, significant CDW correlations and bipolaron formation emerge at strong $e$-ph coupling that suppresses the SC correlations. Previous work\cite{Esterlis, chubukov2020} in the adiabatic limit indeed shows quantitative agreement between ME theory and DQMC in terms of single-particle properties, at least up to a critical value of $\lambda$ beyond which ME theory quickly breaks down. This result demonstrates that while ME theory does not capture the polaronic effects at strong coupling, it does provide an accurate description of the Holstein model's superconducting state in the adiabatic limit.  It is not expected that ME theory will remain valid in the antiadiabatic limit, as vertex corrections proportional to $\lambda \Omega / E_{\text{F}}$ are not included in the theory. In this section, we investigate this expectation and also study the effect of filling on the agreement of ME theory and DQMC.

We compare DQMC to two different versions of ME theory. The unrenormalized version uses a bare phonon propagator for an Einstein mode, whereas the renormalized version self-consistently includes the lowest-order phonon self-energy diagram.\cite{MarsiglioPRB1990} Here, the $\chi_\text{CDW}(\qq)$ is computed within the Migdal approximation by summing the series of particle-hole ring diagrams. Similarly, $\chi_\text{SC}$ is computed by summing the series particle-particle ladder diagrams.

Figure~\ref{migdal} compares our DQMC results with ME theory for various values of the 
phonon frequency. We find that ME theory qualitatively captures the trends in the SC and CDW susceptibilities for weak-coupling and breaks down at larger couplings, consistent with previous studies. Looking more closely at the quantitative agreement, we generally see that renormalized ME theory provides a better agreement wit DQMC with one exception -- the anti-adiabatic limit with $\langle n \rangle=0.8$ suggests that the unrenormalized ME theory does a better job at capturing the SC susceptibility than the unrenormalized ME theory. However, neither renormalized nor unrenormalized ME theory captures both the SC and CDW susceptibilities in this limit. This result is not surprising given ME theory is not excepted to work in the anti-adiabatic limit. 

The bottom panel of Fig.~\ref{migdal} shows the anti-adiabatic limit at low filling ($\langle n \rangle=0.4$). In this case, we observe a surprisingly good agreement between the renormalized ME theory and DQMC up to $\lambda_0 \approx 0.2$. For this limit we have chosen $\Omega / E_{\text{F}} = 10$, which means the breakdown around $\lambda_0 \approx 0.2$ corresponds to a Migdal parameter of $\lambda_0 \Omega / E_{\text{F}} \approx 2$, well outside the regime of validity typically quoted for ME theory. This observation is interesting as it suggests that renormalized Migdal-Eliashberg theory could be applicable as long as the $e$-ph coupling falls within the weak-coupling regime even for materials such as \textit{n}-type $\textrm{SrTiO}_3$, monolayer $\textrm{FeSe}$ on $\textrm{SrTiO}_3$, the fullerides, and lightly doped oxides where the phonon frequency is large compared to the Fermi energy. 

\section{Conclusions}

We have studied the behavior of the superconducting and charge-density-wave susceptibilities across a wide range of parameters for the two-dimensional Holstein model using DQMC, an exact numerical method. The competition between SC, CDW order, and polaronic tendencies is an important aspect of the physics of $e$-ph coupled systems in the intermediate and strong coupling regimes. The strong tendency toward CDW order at half-filling, and the vanishing $T_c$ expected with vanishing carrier concentration, implies that superconductivity is optimized at intermediate carrier densities.  On the other hand, the tendency toward bipolaron formation and/or CDW order for large coupling implies the optimal regime for superconductivity is at intermediate coupling strength. Moreover, we find that larger phonon frequencies favor superconductivity as the CDW correlations are suppressed with increasing $\Omega$ and the behavior of $\chi_\text{SC}$ and $\chi_\text{CDW}(\qq)$ approach that of an effective attractive Hubbard model, as expected in the anti-adiabatic limit~\cite{hubbard}. 
Our results suggest that the maximal superconducting $T_c$ in the Holstein model also increases monotonically with phonon frequency, going as $T_c \approx 0.1\Omega$ in the adiabatic limit and saturating to a value $T_c \approx 0.2 t$ in the anti-adiabatic limit. We stress, however, that the inclusion of the Coulomb interactions leads to significant suppression of $T_c$ unless $\Omega/t$ is very small. 
Finally, we have found that Migdal-Eliashberg theory breaks down at a critical value of $e$-ph coupling $\lambda \sim 1$, regardless of the adiabatic ratio $\Omega/E_{\text{F}}$ or filling, consistent with previous results\cite{Esterlis, chubukov2020, Alexandrov1983, Alexandrov1992, Alexandrov2000, Alexandrov2001, Bauer2011}. However, we find evidence that renormalized Migdal-Eliashberg theory is possibly valid for weak-coupling and low electron densities, even when the phonon frequency exceeds the Fermi energy. This result defies the conventional wisdom that Migdal's theorem is only valid when $\Omega/E_\text{F} < 1$.  \\

\section{ACKNOWLEDGEMENTS}

We thank Douglas Scalapino for his insightful comments and illuminating discussions regarding this work. B.~N., E.~W.~H., B.~M., and T.~P.~D. were supported by the U.S. Department of Energy, Office of Basic Energy Sciences, Division of Materials Sciences and Engineering, under Contract No. DE-AC02-76SF00515. I.~E. and S.~A.~K. were supported by NSF grant \# DMR-1608055 at Stanford. I.~E. acknowledges support from the Harvard Quantum Initiative Postdoctoral Fellowship in Science and Engineering. E.~W.~H. was supported
by the Gordon and Betty Moore Foundation EPiQS Initiative through the grants GBMF 4305 and GBMF 8691. P.~M.~D. and S.~J. were supported by the Scientific Discovery through Advanced Computing (SciDAC) program funded by the U.S. Department of Energy, Office of Science, Advanced Scientific Computing Research and Basic Energy Sciences, Division of Materials Sciences and Engineering. Computational work was performed on the Sherlock computing cluster at Stanford University.


\begin{thebibliography}{9}

\bibitem{CDW} G. Gr\"{u}ner, Rev. Mod. Phys. \textbf{60}, 1129 (1988).

\bibitem{polaron} J. T. Devreese, Polarons, in \textit{Encyclopedia of Applied Physics}, edited by G. L. Trigg (VCH, Weinheim, 1996), Vol. 14, pp. 383-–413.

\bibitem{BCS} J. Bardeen, L. N. Cooper, and J. R. Schrieffer, Phys. Rev. \textbf{108}, 1175 (1957). 

\bibitem{Scalapino} D. J. Scalapino, J. R. Schrieffer, and J. W. Wilkins, Phys. Rev. \textbf{148}, 263 (1966).

\bibitem{Marsiglio} F. Marsiglio and J. P. Carbotte, ``Electron-phonon superconductivity," in Superconductivity: Conventional and Unconventional Superconductors, edited by K. H. Bennemann and J. B. Ketterson (Springer Berlin Heidelberg, Berlin, Heidelberg, 2008) pp. 73-162.

\bibitem{Johnston} S. Johnston, F. Vernay, B. Moritz, Z.-X. Shen, N. Nagaosa, J. Zaanen, and T. P. Devereaux, Phys. Rev. B \textbf{82}, 064513 (2010).

\bibitem{Migdal} A. Migdal, Zh. Eksp. Teor. Fiz. \textbf{34}, 1438 (1958) [Sov. Phys. JETP \textbf{7}, 996 (1958)].

\bibitem{Eliashberg} G. Eliashberg, Zh. Eksp. Teor. Fiz. \textbf{38}, 966 (1960) [Sov. Phys. JETP \textbf{11}, 696 (1960)].

\bibitem{Engelsberg1963}
S. Engelsberg and J. R. Schrieffer, Phys. Rev. \textbf{131}, 993 (1963).

\bibitem{MarsiglioPRB1990}
F. Marsiglio, Phys. Rev. B \textbf{42}, 2416 (1990).

\bibitem{DeePRB2019}
P. M. Dee, K. Nakatsukasa, Y. Wang, and S. Johnston, Phys. Rev. B {\bf 99}, 
024514 (2019).

\bibitem{NosarzewskiPRB2021}
B. Nosarzewski, M. Sch{\"u}ler, and T. P. Devereaux, 
Phys. Rev. B \textbf{103}, 024520 (2021).

\bibitem{LangFirsov} I. G. Lang and Y. A. Firsov, Zh. Eksp. Teor. Fiz. \textbf{43}, 1843 (1962), [Sov. Phys. JETP \textbf{16}, 1301 (1963)]. 

\bibitem{FreericksStrongCoupling} J. K. Freericks, Phys. Rev. B \textbf{48}, 3881 (1993).

\bibitem{Giustino} F. Giustino, Rev. Mod. Phys. \textbf{89}, 015003 (2017).

\bibitem{LiEPL2015}
S. Li and S. Johnston, EPL (Europhysics Letters) {\bf 109}, 27007 (2015).

\bibitem{ScalettarCompetition} R. T. Scalettar, N. E. Bickers, and D. J. Scalapino, Phys. Rev. B \textbf{40}, 197 (1989).

\bibitem{Noack} R. M. Noack, D. J. Scalapino, and R. T. Scalettar, Phys. Rev. Lett. \textbf{66}, 778 (1991).

\bibitem{Vekic} M. Vekic, R. M. Noack, and S. R. White, Phys. Rev. B \textbf{46}, 271 (1992).

\bibitem{Meyer} D. Meyer, A. C. Hewson, and R. Bulla, Phys. Rev. Lett. \textbf{89}, 196401 (2002).

\bibitem{Capone} M. Capone and S. Ciuchi, Phys. Rev. Lett. \textbf{91}, 186405 (2003).

\bibitem{FreericksDMFT} J. K. Freericks, M. Jarrell, and D. J. Scalapino, Phys. Rev. B \textbf{48}, 6302 (1993).

\bibitem{Benedetti} P. Benedetti and R. Zeyher, Phys. Rev. B \textbf{58}, 14320 (1998).

\bibitem{Esterlis} I. Esterlis, B. Nosarzewski, E. W. Huang, B. Moritz, T. P. Devereaux, D. J. Scalapino, and S. A. Kivelson, Phys. Rev. B \textbf{97}, 140501(R) (2018).

\bibitem{LanzaraNature}
A. Lanzara, P. V. Bogdanov, X. J. Zhou, S. A. Kellar, D. L. Feng, E. D. Lu, T. Yoshida, H. Eisaki, A. Fujimori, K. Kishio \textit{et al.}, Nature \textbf{412}, 510-514 (2001).

\bibitem{CukPRL2004}
T. Cuk, F. Baumberger, D. H. Lu, N. Ingle, X. J. Zhou, H. Eisaki, N. Kaneko, Z. Hussain, T. P. Devereaux, N. Nagaosa, and Z.-X. Shen, Phys. Rev. Lett. \textbf{93}, 117003 (2004).

\bibitem{PlumbPRL2010}
N. C. Plumb, T. J. Reber, J. D. Koralek, Z. Sun, J. F. Douglas, Y. Aiura, K. Oka, H. Eisaki, and D. S. Dessau, Phys. Rev. Lett. \textbf{105}, 046402 (2010).

\bibitem{Lee} J. Lee, K. Kujita, K. McElroy, J. A. Slezak, M. Wang, Y. Aiura, H. Bando, M. Ishikado, T. Masui, J.-X. Zhu, A. V. Balatsky, H. Eisaki, S. Uchida, and J. C. Davis, Nature \textbf{442}, 546 (2006).

\bibitem{Devereaux} X. J. Zhou, T. Cuk, T. P.Devereaux, N. Nagaosa, and Z. X. Shen, Handbook of High-Temperature Superconductivity: Theory and Experiment, (New York: Springer) pp. 87–144 (2007).

\bibitem{Zhang} 
A.-M. Zhang and Q.-M. Zhang, Chin. Phys. B \textbf{22}, 087103 (2013).

\bibitem{Farina} D. Farina, G. De Filippis, A. S. Mishchenko, N. Nagaosa, Jhih-An Yang, D. Reznik, Th. Wolf, and V. Cataudella, Phys. Rev. B \textbf{98}, 121104(R).

\bibitem{Pintschovius} L. Pintschovius, phys. stat. sol. (b) \textbf{242}: 30-50 (2005).

\bibitem{CrawfordPRB1990}
M. K. Crawford, M. N. Kunchur, W. E. Farneth, E. M. McCarron III, and S. J. Poon, 
Phys. Rev. B \textbf{41}, 282 (1990).

\bibitem{ChenPNAS2007}
Xiao-Jia Chen, Viktor V. Struzhkin, Zhigang Wu, Hai-Qing Lin, Russell J. Hemley, and Ho-kwang Mao, PNAS \textbf{104}, 3732--3735 (2007). 

\bibitem{Keimer} B. Keimer, S. A. Kivelson, M. R. Norman, S. Uchida and J. Zaanen, Nature \textbf{518}, 179 (2015). 

\bibitem{Chang} J. Chang, E. Blackburn, A. T. Holmes, N. B. Christensen, J. Larsen, J. Mesot, R. Liang, D. A. Bonn, W. N. Hardy, A. Watenphul, M. v. Zimmermann, E. M. Forgan and S. M. Hayden, Nature Phys. \textbf{8}, 871-876 (2012). 

\bibitem{Pablo} Y. Yang, S. Fang, V. Fatemi, J. Ruhman, E. Navarro-Moratalla, K. Watanabe, T. Taniguchi, E. Kaxiras, and P. Jarillo-Herrero, Phys. Rev. B \textbf{98}, 035203 (2018).

\bibitem{Sleight}
A. W. Sleight, Physica C \textbf{514}, 152--165 (2015).

\bibitem{Fisher} D. Nicoletti, E. Casandruc, D. Fu, P. Giraldo-Gallo, I. R. Fisher, and A. Cavalleri, PNAS \textbf{114} (34) 9020-9025 (2017).

\bibitem{Zocco} D. A. Zocco, J. J. Hamlin, K. Grube, J.-H. Chu, H.-H. Kuo, I. R. Fisher, and M. B. Maple, Phys. Rev. B \textbf{91}, 205114 (2015).

\bibitem{LeePreprint2021}
S. Lee, J. Collini, S. X.-L. Sun, M. Mitrano, X. Guo, C. Eckberg, J. Paglione, E. Fradkin, and P. Abbamonte, arXiv:2102.03592 (2021).

\bibitem{StewartReview2015}
G. R. Stewart, Physica C \textbf{514}, 28-35 (2015).

\bibitem{STO} L. P. Gor’kov, PNAS \textbf{113}, 4646 (2016).

\bibitem{FESTO} J. J. Lee, F. T. Schmitt, R. G. Moore, S. Johnston, Y.-T. Cui, W. Li, M. Yi, Z. K. Liu, M. Hashimoto, Y. Zhang, D. H. Lu, T. P. Devereaux, D.-H. Lee, and Z.-X. Shen, Nature \textbf{515}, 245 (2014).

\bibitem{Grimaldi} C. Grimaldi, L. Pietronero, and S. Str{\"a}ssler, Phys. Rev. Lett. \textbf{75}, 1158 (1995).

\bibitem{Gunnarsson} O. Gunnarsson, Rev. Mod. Phys. \textbf{69}, 575 (1997).

\bibitem{Holstein} T. Holstein, Annals of Physics \textbf{8}, 325 (1959).

\bibitem{Dee2020}
P. M. Dee, J. Coulter, K. Kleiner, and S. Johnston, 
Commun. Phys. \textbf{3}, 145 (2020).

\bibitem{owen_scalettar} O. Bradley, G. G. Batrouni, and R. T. Scalettar. arXiv preprint 2011.11703 (2020).

\bibitem{WhiteDQMC} S. R. White, D. J. Scalapino, R. L. Sugar, E. Y. Loh, J. E. Gubernatis, and R. T. Scalettar, Phys. Rev. B \textbf{40}, 506 (1989).

\bibitem{DQMC} S. Johnston, E. A. Nowadnick, Y. F. Kung, B. Moritz, R. T. Scalettar, and T. P. Devereaux, Physical Review B \textbf{87}, 235133 (2013).

\bibitem{esterlis_pseudogap} I. Esterlis, S. A. Kivelson, and D. J. Scalapino, Phys. Rev. B \textbf{99}, 174516 (2019).

\bibitem{Li2020}
S. Li and S. Johnston, npj Quantum Materials \textbf{5}, 40 (2020).

\bibitem{scipy-interpolate}
\url{https://docs.scipy.org/doc/scipy/reference/generated/scipy.interpolate.griddata.html}

\bibitem{allen1982} 
P.B. Allen and B. Mitrovic, \textit{Solid State Physics}, edited by F. Seitz, D. Turnbull and H. Ehrereich (Academic Press, New York, 1982) Vol. 37, pp. 1-92.

\bibitem{hubbard} J. Hubbard, Proc. R. Soc. London Ser. A {\bf 276}, {\it ibid.} 238 (1963); {\it ibid.} 277, {\bf 237} (1964); {\it ibid.} {\bf 281}, 401 (1964); {\it ibid.} {\bf 285}, 542 (1965); {\it ibid.} {\bf 296}, 82 (1967); {\it ibid.} {\bf 296}, 100 (1967).

\bibitem{esterlistc} I. Esterlis, S. A. Kivelson, and D. J. Scalapino, npj Quantum Materials \textbf{3}, 59 (2018).

\bibitem{moreo} A. Moreo and D. J. Scalapino, Phys. Rev. Letters \textbf{66}, 946 (1991).

\bibitem{BergerPRB1995}
E. Berger, P. Val{\'a}{\v s}ek, and W. von der Linden, Phys. Rev. B \textbf{52}, 4806 (1995).

\bibitem{CaponePRB2006}
M. Capone, P. Carta, and S. Ciuchi, Phys. Rev. B \textbf{74}, 045106 (2006).

\bibitem{Bauer}
J. Bauer and A. C. Hewson, Phys. Rev. B \textbf{81}, 235113 (2010).

\bibitem{Weber}
M. Weber and M. Hohenadler, Phys. Rev. B \textbf{98}, 085405 (2018).

\bibitem{Mendl2017}
C. B. Mendl, E. A. Nowadnick, E. W. Huang, S. Johnston, B. Moritz, and T. P. Devereaux
Phys. Rev. B \textbf{96}, 205141 (2017). 

\bibitem{Karakuzu}
S. Karakuzu, L. F. Tocchio, S. Sorella, and F. Becca, 
Phys. Rev. B \textbf{96}, 205145 (2017).

\bibitem{Han2020}
Z. Han, S. A. Kivelson, and H. Yao, Phys. Rev. Lett. \textbf{125}, 167001 (2020)

\bibitem{Costa2020}
N. C. Costa, K. Seki, S. Yunoki, and S. Sorella, Commun. Phys. \textbf{3}, 80 (2020).  

\bibitem{chubukov2020} 
A. V. Chubukov, A. Abanov, I. Esterlis, and S. A. Kivelson, Annals of Physics \textbf{417}, 168190 (2020).

\bibitem{Alexandrov1983} A. S. Alexandrov, Zh. Fiz. Khim. \textbf{57}, 273 (1983); A. S. Alexandrov, Russ. J. Phys. Chem. \textbf{57}, 167 (1983)

\bibitem{Alexandrov1992} A. S. Alexandrov, Phys. Rev. B \textbf{46}, 2838 (1992)

\bibitem{Alexandrov2000} A. S. Alexandrov, Phys. Rev. B \textbf{61}, 12315 (2000)

\bibitem{Alexandrov2001} A. S. Alexandrov, EPL \textbf{56}, 92 (2001)

\bibitem{Bauer2011} J. Bauer, J. E. Han, and O. Gunnarsson, Phys. Rev. B \textbf{84}, 184531 (2011)

\end{thebibliography}
\end{document}